\documentstyle[12pt,epsf,amssymb]{article}

\begin{document}

\baselineskip=18.8pt plus 0.2pt minus 0.1pt

\def\CR{\nonumber \\}
\def\lp{l_P}
\def\pt{\partial}
\def\be{\begin{equation}}
\def\ee{\end{equation}}
\def\bea{\begin{eqnarray}}
\def\bee{\end{eqnarray}}
\def\eq#1{(\ref{#1})}

\begin{titlepage}
\title{
\hfill\parbox{4cm}
{\normalsize KUNS-1630 \\{\tt hep-th/0001161}}\\
\vspace{1cm}
Space-time uncertainty relation and Lorentz invariance}
\author{
Naoki {\sc Sasakura}\thanks{{\tt sasakura@gauge.scphys.kyoto-u.ac.jp}}
\\[7pt]
{\it Department of Physics, Kyoto University, Kyoto 606-8502, Japan}}
\date{\normalsize January, 2000}
\maketitle
\thispagestyle{empty}

\begin{abstract}
\normalsize
We discuss a Lorentz covariant space-time uncertainty relation,
which agrees with that of Karolyhazy-Ng-van\,Dam
when an observational time period
$\delta t$ is larger than the Planck time $\lp$.
At $\delta t\lesssim\lp$, this uncertainty relation takes roughly the
form  $\delta t\delta x\gtrsim \lp^2$, which can be derived from the
condition
prohibiting the multi-production of probes to a geometry.
We show that there exists a minimal area rather than a minimal length in
the four-dimensional case.
We study also a three-dimensional free field theory
on a non-commutative space-time realizing the uncertainty relation.
We derive the algebra among the coordinate and momentum operators
and define a positive-definite norm of the representation space.
In four-dimensional space-time, the Jacobi identity should be violated
in the algebraic representation of the uncertainty relation.
\end{abstract}
\end{titlepage}

\noindent
\section{Introduction}
There is a suspicion that the space-time is not a smooth
manifold in a small scale as assumed in general relativity,
but should be replaced by a new quantum geometry \cite{misner}.
Uncertainty relation of space-time gives
an approach to such quantum space-time.
Since the uncertainty relation of quantum mechanics can be obtained
with a simple thought experiment,
we may expect that the space-time uncertainty relation
might be guessed without knowing the complete theory of quantum gravity.
In fact there are several proposals of the uncertainty relation of
space-time.
Some come from the analysis of the scattering and
the fluctuation sizes of strings and
branes in string theory \cite{scat,yoneya,li,brane}.
Some others are derived from the investigations of the thought measuring
process of space-time distances or locations, using
only the knowledge of the quantum mechanics and the general
relativity \cite{sal,karo,ng,ahl,garay}.
The main idea of the space-time uncertainty relation obtained from such a
thought measuring process is as follows.
To measure the space-time distances or locations precisely, we
need a high energy probe to minimize the space-time size of the wave
function of the probe. But, on the other hand, since gravity couples
to the energy-momentum tensor, the high energy probe would generate large
fluctuations of metric. Since the location of the probe is spread
quantum mechanically, the metric fluctuations have indeterminable
components.
Thus there exists limitation of precise measurement of space-time
distances or locations.

The quantum uncertainty of measuring a space-time distance $l$,
$\delta l \gtrsim (\lp^2 l)^{1/3}$, was obtained first by Karolyhazy
\cite{karo},
and also by Ng and van\,Dam \cite{ng} in a thought experiment first proposed
by
Salecker and Wigner \cite{sal}.
Amelino-Camelia used the uncertainty relation in the form
$\delta l \gtrsim (\lp^2 \delta t)^{1/3}$ with an observational time
period $\delta t$,
and pointed out the possibility of observing this quantum
fluctuation of the geometry with a gravitational-wave
interferometer \cite{amelino}.\footnote{See \cite{adler} for
criticism against this idea, and also \cite{antiadler} for protection}
Barrow argued that this uncertainty relation gives qualitatively the
same prediction of the life-time of a black hole \cite{barrow}
as that derived from Hawking radiation.

In our previous paper \cite{sasa},
we used this uncertainty relation in the form
$\delta V \gtrsim \lp^2 \delta t$, where $\delta V$ denotes an uncertainty
of a spatial volume.
In this case, the uncertainty relation is applied to a
four-dimensional space-time volume.
This interpretation lead to an entropy bound
$S\lesssim\sqrt{EV}/\lp$, where $E$
and $V$ are the total energy and volume of a system, respectively.
We pointed out that this entropy bound fits well to the holographic
principle proposed by 't Hooft and Susskind \cite{hol}.
In fact, a similar kind of entropy bound was assumed
in order to prove rigorously \cite{rig} the Bousso's covariant entropy
bound \cite{bou}.
Recently, a generally covariant formula of a similar entropy bound
was proposed by
Brustein and Veneziano \cite{brusvene},
generalizing the Hubble-entropy-bound \cite{vene}.
Thus it would be a natural expectation that the space-time uncertainty
relation can also be formulated in a generally covariant form.

In this paper, we discuss a Lorentz covariant formulation of the
space-time uncertainty relation above, as the smallest step toward a
generally covariant formulation.
We first point out that the condition that the multi-production
of probe particles should not occur in the measurement process leads to
the uncertainty relation $\delta t \delta x \gtrsim \lp^2$ at
$\delta t\lesssim \lp$. This relation
and $\delta V \gtrsim \lp^2 \delta t$ at $\delta t \gtrsim \lp$
can be naturally combined into the
form $\lp^2 |\delta x_{i_1}^\mu n_\mu| \lesssim
|\varepsilon_{\mu\nu\rho\sigma} n^\mu \delta x_{i_2}^\nu
\delta x_{i_3}^\rho \delta x_{i_4}^\sigma|$ for any observer $n^\mu$.
In general dimensions (larger than two), the uncertainty relation is
just obtained by a straightforward generalization.
With this motivation, we investigate a free field theory
on a non-commutative three-dimensional space-time with the
coordinate commutation relations of Lie algebra $so(2,1)$.
We construct the momentum operators on this non-commutative space-time
and define a positive-definite norm on which the coordinate and
momentum operators are hermite.
We study the Klein-Gordon equation on this non-commutative space-time.
We found a two-fold degeneracy of the spectra.
Lastly we point out that, in the case of the four-dimensional space-time,
the Jacobi identity should be violated in algebraic
representations of the uncertainty relation.

\section{Uncertainty relation of space-time}

The metric tensor field in the general relativity could be
measured by following the trajectory of a point particle.
In classical mechanics, the probe particle can be regarded as a point,
hence we obtain a smooth manifold, which is assumed in the general
relativity.
However, in quantum mechanics, since the probe particle is spread over its
wave function, the measurement becomes necessarily vague.
We shall discuss this measurement uncertainty in a flat Minkowski
space-time.
To be self-contained, we begin with the derivations of the space-time
uncertainty relation of Karolyhazy-Ng-van\,Dam \cite{karo,ng} with a new
ingredient.

We follow, for a time period $\delta t$,
a probe particle with a mass $m$ which has initially
a vanishing total momentum and the gaussian wave packet with
$\langle (\delta x)^2 \rangle_{t=0} = \sigma^2$. We assume the wave function
is symmetric under the spatial rotation. In non-relativistic
approximation, the distribution of the gaussian
wave packet becomes $\langle (\delta x)^2 \rangle_{t=\delta t}=
\sigma^2+ (\delta t)^2/4m^2\sigma^2$
after the time period $\delta t$, by solving the Shr\"odinger
equation. This second term comes essentially from
the momentum distribution, which is inversely proportional to the
spatial distribution $\sigma$. This shows that, even if we
have a smaller wave packet initially, we do not necessarily have a smaller
wave packet for the whole period of the measuring process.
Rather it has a finite minimum
\be
\sigma_{min}\equiv {\rm Min}_{\sigma}
\sqrt{ \sigma^2+ (\delta t)^2/4m^2\sigma^2}.
\label{eq:minsig}
\ee
Without taking into account the condition discussed in the following
paragraph, the minimum value is $\sqrt{\delta t/m}$ at
$\sigma=\sqrt{\delta t/2m}$. At this minimum, the momentum distribution is
in
the order of $\delta p \sim 1/\sigma\sim \sqrt{m/\delta t}$. Hence the
condition for the non-relativistic approximation, $\delta p \lesssim m$, to
hold is equivalent to $m \gtrsim 1/\delta t$. In fact, this condition must
be
imposed from the beginning, since, if not, the
multi-production of the probes will occur because of the quantum
mechanical uncertainty relation between energy and time,
and the measurement by a particle probe must be abandoned.
For a smaller $\sigma$, we should treat in a fully relativistic way.
But the size of the wave packet $\langle (\delta x)^2
\rangle_{t=\delta t}$ in this parameter region
will be anyway larger than the minimum value
$\sigma_{min}$. Thus the non-relativistic approximation is
enough to obtain $\sigma_{min}$ if $m$ is not so near to $1/\delta t$.

The other ingredient of the Karolyhazy-Ng-van\,Dam type uncertainty
relation is the condition that the probe should not become
a black hole for it to be followed.
This condition is hard to evaluate in a reliable manner, since we need
a quantum field theory of general relativity to do so.
We could approximate the condition with that the mean mass distribution
$\rho=m|\varphi(x)|^2$, where $\varphi(x)$ is the gaussian wave function
with
distribution $\sigma$, does not make a
black hole in the classical general relativity, just believing that the
result is qualitatively correct. This condition is given by \cite{wald}
\be
2mG\int_{|x|<R}d^3x |\varphi(x)|^2 < R \ \ \ {\rm for\ all}\ R,
\ee
where $G$ is the gravitational constant. This gives
\be
\sigma \gtrsim \lp^2 m,
\label{eq:condsch}
\ee
which agrees with the intuition that the wave packet cannot be confined
in the size of the Schwarzschild radius of the probe particle.
Thus, from \eq{eq:minsig} and \eq{eq:condsch}, we finally obtain the
shortest observable spatial length in an
observation with a time period $\delta t$:
\be
\delta x_{min}={\rm Min}_{m \gtrsim {1\over \delta t}}{\rm
  Min}_{\sigma \gtrsim \lp^2 m} \sqrt{ \sigma^2+ (\delta t)^2/4m^2\sigma^2},
\label{eq:mindel}
\ee
where we take into account the condition $m \gtrsim {1\over \delta t}$,
which prohibits the multi-production of the probe particles.
This condition is a new ingredient of this paper.
When $\delta t\gtrsim \lp$,
the minimum of \eq{eq:mindel} is at $\sigma=\lp^2 m,\
m^6=(\delta t)^2/2\lp^8$.
Thus we obtain the Karolyhazy-Ng-van\,Dam type uncertainty relation
\cite{karo,ng,amelino}:
\be
(\delta x)^3\gtrsim (\delta x_{min})^3 \sim \lp^2 \delta t.
\label{eq:x3}
\ee
The non-relativistic approximation is valid for this case, since
$m/\delta p\sim \sigma m \sim  (\delta t/\lp)^{2/3}\gtrsim 1$ at the
minimum.
When $\delta t \lesssim \lp$, the minimum is at $\sigma=\lp^2 m,\
m=1/\delta t$. Thus we obtain
\be
\delta x \gtrsim \delta x_{min}\sim {\lp^2 \over \delta t}.
\label{eq:xt}
\ee
Again the non-relativistic approximation is valid, since
$m/\delta p \sim \sigma m \sim (\lp/\delta t)^2\gtrsim 1$ at the
minimum. Form the inequalities \eq{eq:x3} and \eq{eq:xt}, one founds that
the Planck length is the minimal spatial length, $\delta x \gtrsim
\lp$, while $\delta t$ is not bounded. This rather odd result will be
remedied properly in the Lorentz covariant formulation discussed below.

The relations \eq{eq:x3} and \eq{eq:xt} can be rewritten in the following
suggestive ways:
\bea
(\delta x)^3 &\gtrsim & \lp^2 \delta t, \CR
(\delta x)^2 \delta t &\gtrsim & \lp^2 \delta x.
\label{eq:sgt}
\bee
Thus we propose that the uncertainty relation of space-time
has the following Lorentz covariant form:
\be
\lp^2 \delta x_{i_1}^\mu n_\mu\lesssim
|\varepsilon_{\mu\nu\rho\sigma} n^\mu \delta x_{i_2}^\nu
\delta x_{i_3}^\rho \delta x_{i_4}^\sigma| \ \ {\rm for\ any}\ n^\mu,
\label{eq:lorine}
\ee
where $\delta x_i^\mu(i=1,2,3,4)$ are the four vectors defining a space-time
volume, and the inequality \eq{eq:lorine} should be satisfied for
all ordering of $\delta x_i^\mu$. The case $\delta x_1=(\delta t,0,0,0),
\delta x_2=(0,\delta x,0,0),\delta x_3=(0,0,\delta x,0),
\delta x_4=(0,0,0,\delta x)$ in \eq{eq:lorine} reproduces \eq{eq:sgt}.
By substituting $n^\mu=\delta x_i^\mu$ in \eq{eq:lorine}, we see that
the four vectors $\delta x_i^\mu$ must be orthogonal among each other.

The condition \eq{eq:lorine} might be too strong. Physically, the
vector $n^\mu$ would denote the velocity vector of an observer.
In fact, the following weaker condition is enough to reproduce \eq{eq:sgt}:
\be
\lp^2 |\delta x_{i_1}^\mu n_\mu| \lesssim
|\varepsilon_{\mu\nu\rho\sigma} n^\mu \delta x_{i_2}^\nu
\delta x_{i_3}^\rho \delta x_{i_4}^\sigma| \ \
{\rm for\ any\ time\mbox{-}like}\ n^\mu.
\label{eq:lorinetime}
\ee
There would be other possible Lorentz covariant formulations which
reproduce \eq{eq:sgt}. Presently, we do not have any criteria to choose
one of them.

Now let us consider the case
$\delta x_1=(\delta t,0,0,0), \delta x_2=
(0,\delta l_1,0,0),\delta x_3=(0,0,\delta l_2,0),\delta
x_4=(0,0,0,\delta l_3)$.
{}From \eq{eq:lorine} or \eq{eq:lorinetime}, we obtain
\bea
\lp^2 \delta t \lesssim \delta l_1 \delta l_2 \delta l_3, \CR
\lp^2 \delta l_1 \lesssim \delta l_2 \delta l_3 \delta t, \CR
\lp^2 \delta l_2 \lesssim \delta l_3 \delta l_1 \delta t, \CR
\lp^2 \delta l_3 \lesssim \delta l_1 \delta l_2 \delta t.
\bee
{}From the first and the second equation, we find
\be
\delta l_2 \delta l_3 \gtrsim \frac12 \lp^2 \left(
\frac{\delta l_1}{\delta t}+\frac{\delta t}{\delta l_1} \right)
\gtrsim \lp^2.
\ee
Similarly, we obtain
\bea
&&\delta l_1 \delta l_2 \gtrsim \lp^2, \CR
&&\ \ \ \vdots\ \ \ \ \ .
\bee
Thus we conclude that the area is bounded from below by the square of
the Planck length, while the length is not in general. We have found
a minimal area rather than a minimal length in four dimensional space-time.

The generalization of \eq{eq:lorine} and \eq{eq:lorinetime}
to any space-time dimension is obvious.
Let us consider a three-dimensional case with
$\delta x_1=(\delta t,0,0), \delta x_2=
(0,\delta l_1,0),\delta x_3=(0,0,\delta l_2)$.
The uncertainty relation leads to
\bea
\lp \delta t \lesssim \delta l_1 \delta l_2, \CR
\lp \delta l_1 \lesssim \delta l_2 \delta t, \CR
\lp \delta l_2 \lesssim \delta l_1 \delta t.
\label{eq:threeunc}
\bee
{}From these inequalities, we obtain
$\delta t, \delta l_1, \delta l_2 \gtrsim \lp$.
Thus there is a minimal length in three-dimensional space-time.

\section{Algebraic representation}

In this section, we shall discuss an algebraic representation of
the uncertainty relation \eq{eq:threeunc}
in three-dimensional space-time in the spirit
of Snyder \cite{sny}, and study a free field theory on a
non-commutative space-time obtained from it.
Lastly we give some comments for the four-dimensional case.

The algebra we use for the coordinates is the Lie algebra $so(2,1)$:
\be
[x^\mu,x^\nu]=i\lp\varepsilon^{\mu\nu\rho}x_\rho.
\label{eq:so21}
\ee
The motivation to use this algebra is that, if there exists a
semi-positive definite norm on the representation space of $x^\mu$
with an appropriate $\dagger$ operation,
we can show that
\bea
&&\lp^2 \langle x^0 \rangle^2
\leq 4\langle (x^1)^2 \rangle \langle (x^2)^2 \rangle,\CR
&&\ \ \ \vdots\ \ \ ,
\label{eq:uncthr}
\bee
by using $\langle (x^1+i\lambda x^2)^\dagger (x^1+i\lambda x^2)\rangle
\geq 0$ and so on for any $\lambda$. The inequalities \eq{eq:uncthr} may be
regarded as a realization of
the space-time uncertainty relation \eq{eq:threeunc}.

Firstly we shall discuss the momentum operator on this space-time.
Let us assume the following form of the commutation relations for the
momentum operators:
\bea
 \mbox{[} p^\mu,p^\nu \mbox{]} &=&0, \CR
 \mbox{[} p^\mu,x^\nu \mbox{]} &=&-i \eta^{\mu\nu} f(\lp^2 p^2) + i\lp
\varepsilon^{\mu\nu\rho}p_\rho g(\lp^2 p^2),
\label{eq:comp}
\bee
where $f,g$ are some functions with $f(0)=1$ so that we recover the
usual commutation relations in the limit $\lp\rightarrow0$.
In \eq{eq:comp}, we have ignored a term with the tensor structure
$p^\mu p^\nu$, because this term could be absorbed by the redefinition
$p^\mu\rightarrow p^\mu k(\lp^2p^2)$.
We take the signature
$\eta^{\mu\nu}=(-1,1,1)$ and $\varepsilon^{012}=1$.
The criteria to determine these functions $f,g$ are the Jacobi identities.
For the cases $[x,[x,x]+\cdots$, $[x,[p,p]]+\cdots$ and
$[p,[p,p]]+\cdots$, the Jacobi identities are satisfied obviously.
In the remaining case, we have
\bea
&&\mbox{[}p^\mu,\mbox{[}x^\nu,x^\rho\mbox{]}\mbox{]}+
\mbox{[}x^\rho,\mbox{[}p^\mu,x^\nu\mbox{]}\mbox{]}+
\mbox{[}x^\nu,\mbox{[}x^\rho,p^\mu\mbox{]}\mbox{]} \CR
&&\ \ \ \ =\lp \varepsilon^{\mu\nu\rho} f(1-2g)
+\lp^2 (\eta^{\mu\nu} p^\rho-\eta^{\mu\rho}p^\nu)(2ff'-g+g^2)\CR
&&\ \ \ \ \ \ \ +2\lp^3 (\varepsilon^{\mu\rho\sigma}p_\sigma p^\nu-
\varepsilon^{\mu\nu\sigma} p_\sigma p^\rho) g'f.
\bee
For this to vanish, $f$ and $g$ are determined as
\be
f(u)=\sqrt{1+\frac14u},\ \ \ g(u)=\frac12.
\label{eq:fg}
\ee

The algebra \eq{eq:comp} and \eq{eq:fg} we have obtained has a great
similarity with that discussed by Maggiore \cite{mag}. From
\eq{eq:comp} and \eq{eq:fg}, we immediately obtain
\be
\Delta x^\mu \Delta p^\mu \gtrsim \frac12
\left\langle \sqrt{1+\frac{\lp^2 p^2}{4}} \right\rangle,
\label{eq:uncpq}
\ee
where the summation over $\mu$ is not assumed.
As was discussed by Maggiore, this uncertainty relation \eq{eq:uncpq} has
the
following interesting features. In the high Euclidean momentum
regime with $\langle p^i \rangle^2\sim (\Delta p^i)^2 \gg \lp^2$, we get
\be
\Delta x^i \gtrsim \frac{\lp}{4}.
\ee
Thus we obtain a minimal spatial length.
However, since, in our Lorentzian case,
the right-hand side of \eq{eq:uncpq} can take a fixed
value by canceling a  spatial momentum with its time-component,
we do not argue that there exists a universal minimal
length in the sense discussed in \cite{kempf}.
In the regime $|\langle p^2 \rangle| \ll \lp^{-2}$, we obtain
an uncertainty relation similar to that from string theory
\cite{scat,yoneya}:
\be
\Delta x^\mu \Delta p^\mu \gtrsim \frac12+ \frac{\lp^2
\langle p^2 \rangle}{16}.
\ee

Now we shall discuss an explicit representation of the operators which
satisfy the commutation relations \eq{eq:so21} and \eq{eq:comp} with
\eq{eq:fg}.
Since the momentum operators are commutative, we work in the
representation in which the momentum operators are diagonal.
For $x^\mu$ to satisfy \eq{eq:so21} and \eq{eq:comp} with \eq{eq:fg}, we
easily obtain
\be
x^\mu=-\frac{i\lp}2 \varepsilon^{\mu\nu\rho}p_\nu\frac{\pt}{\pt p^\rho}+
i\sqrt{1+\frac{\lp^2 p^2}4}\frac{\pt}{\pt p_\mu}+ \lp^2 A(\lp^2p^2)p^\mu,
\label{eq:opx}
\ee
where $A(\lp^2p^2)$ is an arbitrary function of $p^2$.
The measure of the representation space is also determined uniquely
up to a multiplication constant under the condition that the operators
$x^\mu$ and $p^\mu$ be hermite. It is
\be
\langle \Psi_1 | {\cal O}(x,p) |\Psi_2\rangle
=\int d^3p \left(1+\frac{\lp^2 p^2}4 \right)^{-\frac12}
\Psi_1^*(p) {\cal O}(x,p) \Psi_2(p).
\label{eq:measure}
\ee
For $x^\mu$ to be hermite, the arbitrary function $A(\lp^2 p^2)$ of
\eq{eq:opx} must be a real function.

The measure \eq{eq:measure} sets bounds on the range of the momentum
$p^\mu$:
\be
- p^2 < \frac4{\lp^2}.
\label{eq:boundp}
\ee
Thus there exists a mass upper-bound.
{}From the explicit representation \eq{eq:opx} of $x^\mu$, we can
see that the operation $x^\mu$ does not violate this bound. This is
obvious for the first and third terms of \eq{eq:opx}. The second
term changes $p^2$ in general, but, at the boundary
$p^2=-\frac4{\lp^2}$, the change vanishes.
However, as we will see soon, there is a little complication
concerning the global structure of the representation space, and
we will find that the momentum representation space is in fact
two-fold degenerate.

Before discussing this feature,
let us look for a wave function $\Psi$ which satisfies the
Klein-Gordon equation, $(p^\mu p_\mu+m^2)
\Psi=0$. The solution is trivially given by
$\Psi_{\pm}(p)=\theta(\pm p^0)\delta(p^2+m^2)$
in the momentum representation. There is also another route to obtain the
wave
function in terms of the coordinates $x^\mu$.
Let us assume the following form of the wave function:
\be
\Psi(k^\mu)=e^{ik^\mu x_\mu}|0\rangle,
\ee
where $k^\mu$ are $c$-numbers, and
$|0\rangle$ denotes the momentum zero eigenstate $p^\mu|0\rangle=0$.
Now we operate $p^\mu$ on this state:
\bea
p^\mu e^{ik^\mu x_\mu}|0\rangle &=&  \int_0^1 ds e^{i(1-s)k^\mu x_\mu}
\mbox{[}p^\mu,ik^\nu x_\nu\mbox{]} e^{isk^\mu x_\mu} |0\rangle \CR
&=&\int_0^1 ds e^{i(1-s)k^\mu x_\mu}
\left( 
k^\mu\sqrt{1+\frac14\lp^2p^2}-\frac12\lp\varepsilon^{\mu\nu\rho}
k_\nu p_\rho \right) e^{isk^\mu x_\mu} |0\rangle 
\label{eq:pfirst}
\bee
where we have used the formula $\frac{d}{dt} e^A=\int_0^1 ds e^{(1-s)A}
\frac{dA}{dt} e^{sA}$ and the commutation relations \eq{eq:comp}.
Repeating the above operation of $p^\mu$ perturbatively in $\lp$, 
one can see that the wave
function $\Psi(k^\mu)$ is in fact an eigenstate of the momentum operator
$p^\mu$:
\be
p^\mu e^{ik^\mu x_\mu}|0\rangle = k^\mu h(k^2) e^{ik^\mu x_\mu}
|0\rangle.
\label{eq:peigen}
\ee
To determine the function $h(u)$, we substitute \eq{eq:peigen} into 
\eq{eq:pfirst}.
Then we obtain an integration equation,
\be
h(u)=\int_0^1 ds \sqrt{1+\frac14 \lp^2  (h(s^2u))^2 s^2 u}.
\ee
By taking the derivative with respect to $u$, we obtain a differential 
equation
\be
2uh'+h-\sqrt{1+\frac14\lp^2 u h^2}=0.
\ee
The solution of this equation is given by
\bea
h(k^2)&=&\frac2\lp \frac1{\sqrt{k^2}}{\rm sinh}\left(\frac\lp{2}
  \sqrt{k^2}\right) \ \ {\rm for\ a\ space\mbox{-}like}\ k^2, \CR
h(k^2)&=&\frac2\lp \frac1{\sqrt{-k^2}}{\rm sin}\left(\frac\lp{2}
  \sqrt{-k^2}\right) \ \ {\rm for\ a\ time\mbox{-}like}\ k^2. 
\bee
Thus the field equation becomes
\be
(p^\mu p_\mu +m^2)\Psi(k^\mu)=\left(-\frac4{\lp^2}{\rm sin}^2\left(\frac\lp{2}
  \sqrt{-k^2}\right) +m^2\right) \Psi(k^\mu)=0.
\label{eq:fieldeq}
\ee
In the case $m=0$, the solutions for \eq{eq:fieldeq} are:
\be
k^2=-\left(\frac{2\pi n}{\lp}\right)^2
\ee
with integer $n$. 
Thus it looks apparently 
as if there are an infinite number of additional states.

To study this issue in more detail, let us consider the operator 
$\Omega=e^{2\pi i x^0/\lp}$. The following discussion is obviously
applicable also to the Lorentz transform of $\Omega$.
Th
e operator $\Omega$ commutes with the coordinate $x^\mu$, because it
is just the $2\pi$ rotation in the spatial plane.
To see the operation on the momentum $p^\mu$, let us consider the
operator $\Omega(s)=e^{is x^0/\lp}$ with an arbitrary real number
$s$, and define $p^\mu(s)=\Omega(s)p^\mu\Omega(-s)$. Then we have
\bea
\frac{dp^\mu(s)}{ds}&=&\frac{i}{\lp} [x^0,p^\mu(s)] \CR
&=& \left\{
\begin{array}{l}
 \frac{1}{\lp}\sqrt{1+\frac{\lp^2p^2(s)}{4}}\ \ {\rm for}\ \mu=0, \\
 \frac12 \varepsilon^{ij}p_j(s)\ \ {\rm for}\ \mu=i,
\end{array}
\right.
\label{eq:diffp}
\bee
where $i,j$ denote the indices for the two-dimensional spatial plane,
and $\varepsilon^{ij}$ is the antisymmetric tensor with
$\varepsilon^{12}=1$.
The differential equation \eq{eq:diffp} is easily solved with the
parameterization $p^0(s)=(2/\lp)\sqrt{1+\lp^2a^2/4}\sin\theta,
p^1(s)=a\cos\varphi,p^2(s)=a\sin\varphi$:
\be
\theta-\theta_0=\varphi-\varphi_0=-\frac12 s,
\label{eq:solves}
\ee
where $\theta_0$ and $\varphi_0$ are the initial values.
Thus $\Omega=\Omega(2\pi)$ is a PT transform of the states.
By performing the operation $\Omega$ twice,
a state will be transformed to the identical state.
However, as we will explain in the followings, a state
obtained by operating $\Omega$ once is not identical with
the anti-particle state of the original state.
In the derivation of \eq{eq:solves},
we assumed that the square root in \eq{eq:diffp} or
\eq{eq:fg} can take both positive and negative values:
$\sqrt{1+\lp^2p^2(s)/4}=\sqrt{1+a^2\lp^2/4}\cos\theta$.
In other words, there are two patches where the square root takes
positive or negative values.
Thus it is more appropriate to
parameterize the representation space in terms of $\theta$ and $p^i$ with
$p^0=(2/\lp)\sqrt{1+(p^i)^2\lp^2/4}\sin\theta$, where $\theta$ has
the range $0\leq\theta<2\pi$ with the identification between $\theta=0$ and
$\theta=2\pi$.
With this parameterization, the measure becomes
\be
\int d^3p (1+p^2\lp^2/4)^{-\frac12}=\frac2\lp \int d\theta d^2p.
\ee
Thus we have obtained safely a positive definite norm.
Since, for a given value of momentum, we have two choices for
$\theta$ (unless $\theta=\pm\pi/2$), we conclude that
the momentum representation space
as well as the spectra of the states satisfying the Klein-Gordon
equation are two-fold degenerate.

In \eq{eq:so21}, the translational symmetry is not manifest.
As was discussed in \cite{sny}, the notion of translational symmetry
should be appropriately substituted for the non-commutative
space-time. The simplest candidate for the translation by a c-number
vector $v^\mu$ is the multiplication of $e^{-iv^\mu p_\mu}$ on a wave
function $\Psi(p^\mu)$ in the momentum representation.
In fact, from \eq{eq:opx}, we obtain $e^{iv^\mu p_\mu}x^\mu
e^{-iv^\mu p_\mu}=x^\mu+k^\mu$ in the limit $\lp\rightarrow 0$.
The whole theory should respect this symmetry.

Finally let us discuss whether it is possible to represent the
uncertainty relation algebraically in four dimensions
as in the three-dimensional case.
The uncertainty relation has roughly the form
$\varepsilon_{\mu\nu\rho\sigma}\delta x^\mu \delta x^\nu \delta x^\rho
\gtrsim \lp^2 \delta x_\sigma$. We now assume the greater side is
represented as a commutation of operators.  Moving one of the
coordinate operator from the left to the right-hand side as the
momentum operator, a possibility is
\be
\mbox{[}x^\mu,x^\nu\mbox{]}=i \lp^2 \varepsilon^{\mu\nu\rho\sigma}
p_\rho x_\sigma+ o(\lp^2).
\label{eq:our4}
\ee
This algebra has a similarity with the Snyder algebra \cite{sny}.
The Snyder algebra has the form $[x^\mu,x^\nu]=ia^2
J^{\mu\nu}$, where $a$ and $J^{\mu\nu}$ denote the length scale
and the Lorentz transformation generators associated to the non-commutative
space-time, respectively. The difference is that, in our case \eq{eq:our4},
the right-hand side is twisted:
\be
[x^\mu,x^\nu]\sim i \lp^2 \varepsilon^{\mu\nu\rho\sigma} J_{\rho\sigma}.
\ee
However we cannot proceed further to obtain the full set of
the commutation relations as in the three-dimensional case.
This is because, using $[p^\mu,x^\nu]=-i\eta^{\mu\nu}+o(\lp)$, we obtain
\be
[x^\mu,[x^\nu,x^\rho]]=-\lp^2 \varepsilon^{\mu\nu\rho\sigma}
x_\sigma+o(\lp^2),
\ee
and the Jacobi identity should be violated in the four-dimensional case.

Another possibility of representing the uncertainty relation is
\be
[x^\mu,p^\nu]=i\eta^{\mu\nu}+\lp^2\varepsilon^{\mu\nu\rho\sigma}p_\rho
p_\sigma.
\ee
However, since $[p^\mu,p^\nu]=o(1)$, the last term would be smaller
than the order of $\lp^2$, and can not work as a realization of the
uncertainty relation.

The violation of the Jacobi identity is not a special thing
mathematically, and is usual when the algebra is not
associative \cite{okubo}.
However, the associativity seems to be deeply embedded
in the formulation of quantum mechanics. Before going further,
we might have to know
the physical reason why we should abandon the associativity.
Without any physical motivations, we would have another possibility using
some triple product $[a,b,c]$ to represent the uncertainty relation
in a form
$\varepsilon_{\mu\nu\rho\sigma}[x^\mu,x^\nu,x^\rho]=\lp^2x_\sigma$.

\section{Summary and discussions}

In this paper, we have investigated a Lorentz covariant
formulation of a space-time uncertainty relation.
The two uncertainty relations at $\delta t \gtrsim \lp$ and
$\delta t\lesssim\lp$ derived from a thought experiment
are combined into a single inequality.
We find a minimal area in the four-dimensional case, while we find a
minimal length in three dimensions.
This can be obviously generalized to any space-time dimension $D$, in
which a $(D-2)$-volume has a lower bound.

We also discussed a free field theory on a non-commutative space-time,
which realizes the space-time uncertainty relation above.
We have obtained a representation of the coordinate operators and
a positive-definite norm in the momentum representation.
The momentum space is two-fold degenerate, and a further investigation
would be needed to clarify its origin.
We have shown that there exists a mass upper-bound.
The existence of the mass upper-bound is an expected property of
three-dimensional gravity.
It is known that
a point-particle generates a conical singularity with a deficit angle
proportional to its mass \cite{deser}.
Since a deficit angle cannot exceed
$2\pi$, a mass should have an upper bound.
In four dimensions, we seem to be forced to use an unusual algebra
to represent the uncertainty relation.

Our space-time uncertainty relation does not respect the parity invariance.
Since this invariance is not respected also in the standard model, the
violation at the fundamental geometrical level would
be interesting rather than disappointing.

The three-dimensional non-commutative space-time seems to be
consistently formulated.
It would be highly interesting to construct a quantum field theory on
this non-commutative space-time and investigate its thermodynamics to
compare with the intuitive arguments in our previous paper \cite{sasa}.
We hope we can give some results in this direction in our future works.

\vspace{.5cm}
\noindent
{\large\bf Acknowledgments}\\[.2cm]
The author would like to thank T.~Yoneya and H.~Suzuki for
valuable suggestions, and D.V.~Ahluwalia, A.~Kempf, M.~Li and Y.J.~Ng
for interesting communication. He is also grateful for the hospitality
at Summer Institute '99, Yamanashi, Japan, where this work was
initiated.
He was supported in part by Grant-in-Aid for Scientific Research
(\#09640346), and in part by
Priority Area: ``Supersymmetry and Unified Theory of Elementary
Particles'' (\#707), from Ministry of Education, Science, Sports and
Culture.

\newcommand{\J}[4]{{\sl #1} {\bf #2} (#3) #4}
\newcommand{\andJ}[3]{{\bf #1} (#2) #3}
\newcommand{\AP}{Ann.\ Phys.\ (N.Y.)}
\newcommand{\MPL}{Mod.\ Phys.\ Lett.}
\newcommand{\NP}{Nucl.\ Phys.}
\newcommand{\PL}{Phys.\ Lett.}
\newcommand{\PR}{Phys.\ Rev.}
\newcommand{\PRL}{Phys.\ Rev.\ Lett.}
\newcommand{\PTP}{Prog.\ Theor.\ Phys.}
\newcommand{\JMP}{J.\ Math.\ Phys.}

\end{document}